# A novel solution for seepage problems using physics-informed neural networks

Tianfu Luo[1], Yelin Feng[2], Qingfu Huang[2], Zongliang Zhang[2], Mingjiao Yan[1], Zaihong Yang[2], Dawei Zheng[2], Yang Yang[2*]

1 College of Water Conservancy and Hydropower Engineering, Hohai University, Nanjing 210098, China;

2 PowerChina Kunming Engineering Corporation Limited, Kunming 650051, China;

* Correspondence: Yang Yang (e-mail:yangyang_kmy@powerchina.cn)

**Abstract:** A Physics-Informed Neural Network (PINN) provides a distinct advantage by synergizing neural networks' capabilities with the problem's governing physical laws. In this study, we introduce an innovative approach for solving seepage problems by utilizing the PINN, harnessing the capabilities of Deep Neural Networks (DNNs) to approximate hydraulic head distributions in seepage analysis. To effectively train the PINN model, we introduce a comprehensive loss function comprising three components: one for evaluating differential operators, another for assessing boundary conditions, and a third for appraising initial conditions. The validation of the PINN involves solving four benchmark seepage problems. The results unequivocally demonstrate the exceptional accuracy of the PINN in solving seepage problems, surpassing the accuracy of FEM in addressing both steady-state and free-surface seepage problems. Hence, the presented approach highlights the robustness of the PINN and underscores its precision in effectively addressing a spectrum of seepage challenges. This amalgamation enables the derivation of accurate solutions, overcoming limitations inherent in conventional methods such as mesh generation and adaptability to complex geometries.

## 1 Introduction

Seepage is the flow of a fluid in a porous medium, and it is a crucial issue for engineering. Seepage problems will generally focus on analyzing the distribution of the hydraulic head [1], the pore pressure [2], the local hydraulic gradient [3], the flow rate [4], and other fields [5–7]. Only a few simple cases can be solved analytically [8, 9]; most are approximate analytical solutions [10], but it is not easy to use directly to analyze practical situations. Physical tests can obtain valuable data [11, 12]. Still, factors such as observation techniques [13, 14] and size effects [15] can limit the application of physical tests to complex engineering seepage problems. Therefore, with the rapid development of computer science and numerical solutions of various types of partial differential equations (PDEs), numerical analysis has gradually become one of the essential tools in complex engineering problems. By harnessing the power of numerical methods, engineers and researchers can gain precise and comprehensive insights into the intricate dynamics of fluid flow in practical scenarios. The numerical analysis offers several notable advantages over alternative approaches, as it enables the accurate representation of complex geometries and boundary conditions [16, 17], thereby facilitating a more realistic portrayal of real-world situations. The percolation inhomogeneous anisotropic porous media equation can be considered a diffusion equation, also known as a heat conduction-type equation.

Numerical solutions of the seepage equation involve various methods, such as the finite volume method (FVM) [18–20], the finite element method (FEM) [21, 22], the finite difference method (FDM) [23–25], the finite analysis method (FAM) [26], the meshless method [27–29], the boundary element method (BEM) [30–32], the scaled boundary finite element method (SBFEM) [33], and more. Among these, the FEM and the FDM are mesh-dependent methods that require the generation of a boundary-fitted mesh. Consequently, mesh quality plays a significant role in the accuracy and reliability of the results obtained using these methods. The dependence on mesh quality can be considered a limitation of the FEM and the FDM. Mesh generation can be complex and time-consuming, especially for intricate geometries. It requires expertise and careful consideration to ensure an appropriate mesh that accurately representing the underlying domain. It requires expertise and careful consideration to ensure an appropriate mesh accurately representing the underlying domain. The success of the FEM and the FDM is closely tied to the suitability and refinement of the mesh. Determining the seepage free-surface often relies on the finite element meshing method. As a part of the seepage boundary, its position is unknown in advance and changes with the change of the upstream water level. When traditional methods simulate complex boundaries, The seepage problem faces problems of low accuracy and grid dependence [34]. Methods such as the BEM and the meshless method alleviate the need for a boundary-fitted mesh, simplifying the preprocessing stage considerably [35]. The meshless method has significant advantages as it reduces the complexity and computational effort required to generate the mesh and is more adaptable to seepage free-surface problems [35]. The BEM, in particular, relies on discretizing only the domain's boundaries, making it suitable for problems with infinite domains or those primarily influenced by boundary conditions. Yang et al. [33, 36] solved the heat conduction problem and seepage problem through a new method that combines SBFEM with polygonal mesh technology.

Deep neural networks (DNNs) are machine learning models that have shown success in many different problems, such as computer vision (CV) [37], natural language processing (NLP) [38], automated driving [39], etc. DNNs are used in artificial intelligence and have made progress and interest in engineering and scientific computing. For example, machine learning has made some progress in earthquake monitoring [40, 41] and fluid simulation [42, 43]. The development of open-source machine learning platforms such as TensorFlow [44] and PyTorch [45] has enabled the rapid growth of deep neural networks in scientific and engineering computing [46]. Deep learning is a sub-learning of machine learning that learns multi-layer features from data and makes predictions. Currently, models for deep learning mainly use neural network models.

It is essential to consider an alternative approach that addresses the limitations of mesh-dependent methods while preserving their strengths. In recent years, the emergence of Physics-Informed Neural Networks (PINN) has gained attention as a promising methodology for solving the PDEs [47, 48]. The PINN offers several advantages in comparison to traditional methods. Firstly, the PINN does not require a predefined mesh, which eliminates the need for mesh generation and simplifies the preprocessing step significantly. This feature allows for more straightforward implementation and reduces the computational overhead associated with mesh refinement and adaptation. Secondly, the PINN combines the power of neural networks with the underlying physics of the problem. The PINN ensures that the solutions satisfy the fundamental laws of physics throughout the domain by encoding the governing equations into the network architecture. This physics-informed training process enables the PINN to capture complex spatiotemporal patterns and accurately simulate the system's dynamics. While the FEM and the FDM have become established methods for solving the PDEs, the advent of PINN represents a significant advance. The PINN provides a mesh-free approach that circumvents the limitations associated with mesh generation and allows greater flexibility when dealing with complex problems. By combining the advantages of traditional methods with the capabilities of neural networks, the PINN provides a promising alternative for efficient and accurate PDEs solutions in various scientific and engineering applications.

Less research into PINN solutions for seepage problems has been carried out. The existing research on PINN solutions to infiltration problems is focused on data-driven [49] or hybrid data and physical constraint-driven approaches [50, 51]. Data-driven methods require large amounts of data to train the model and have drawbacks such as poor generalization [52]. Physical constraint-driven has higher generalization and predictive power compared to data-driven. This paper uses physical constraints to drive physical neural networks to solve various infiltration problems, including steady-state seepage problems, transient seepage problems, and free-surface seepage problems.

This study addresses seepage challenges encompassing steady-state seepage, saturated transient seepage, and free-surface seepage utilizing PINN. The paper unfolds in six segments: Section 2 delineates the governing equations governing seepage. Section 3 provides an overview of the resolution process. Section 4 elucidates the tailored PINN designed to tackle percolation issues. Subsequently, Section 5 showcases various benchmark examples. Ultimately, Section 6 encapsulates the key conclusions drawn from this investigation.

## 2 Methodology

2.1 Governing equation of seepage problems

The governing differential equation for two-dimensional seepage problem can be written as [53]

$$k_x \frac{\partial^2 h}{\partial x^2} + k_y \frac{\partial^2 h}{\partial y^2} = S_s \frac{\partial h}{\partial t}, \qquad (1)$$

where $k_x$ and $k_y$ are the components of the hydraulic conductivity in the horizontal and vertical directions, respectively; $h$ is the hydraulic head; $S_s$ is the specific storage; $t$ is time. In order to address differential equations, it is imperative to prescribe boundary conditions. The Dirichlet boundary condition pertains to specifying the variable sought on the boundary. In contrast, the Neumann boundary condition specifies the gradient of the variable sought at the given boundary. Consequently, the boundary conditions may be expressed as follows:

$$h(x, y) = e \quad \text{on} \quad \Gamma_D, \qquad (2)$$

$$\frac{\partial h(x, y)}{\partial n} = f \quad \text{on} \quad \Gamma_N, \qquad (3)$$

where $e$ and $f$ are the values of the Dirichlet and Neumann boundary, respectively. $\Gamma_D$ and $\Gamma_f$ denote the Dirichlet and Neumann boundary conditions, respectively.

## 2.2 Neural networks and deep learning

The neural network represents a machine learning model characterized as a network or circuitry of neurons. In a contemporary context, it manifests as an artificial neural network comprised of interconnected nodes or neurons [51]. The neural network serves as a computational model designed to emulate the neural system of the human brain, incorporating numerous interconnected neurons that facilitate data transmission. Weight coefficients are associated with the connections between individual neurons, and these weights differ across distinct nodes. Incoming data from connected nodes are weighted and subjected to computational operations. These weights serve to modulate the influence of input data, while biases are employed to control the ease of neuron activation.

Additionally, the activation function determines the aggregate of input data. Illustrated in Figure 1, the entire data originating from the input neuron undergoes multiplication by the weight, followed by the addition of the bias, ultimately yielding the output of the neuron post-application of the activation function. Furthermore, this process can be formally expressed as Equation (4) and Equation (5).

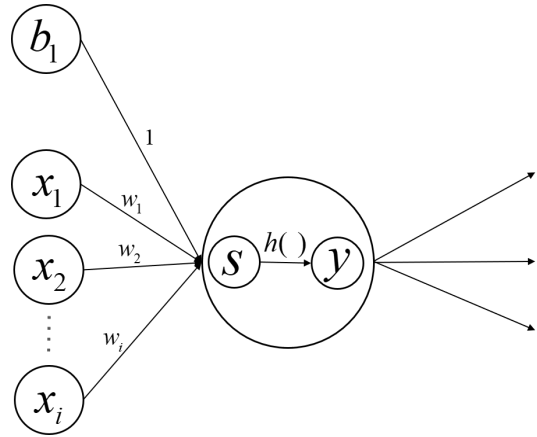

Figure 1. A diagram of the computational process of a neuron, where $x_1, x_2, \ldots, x_i$ are the input data to the neuron; $w_1, w_2, \ldots, w_i$ are the weights corresponding to the different input data; $b_1$ is the bias; $S$ is the input data is multiplied by the weight plus the bias, calculated according to Eq. (4); $h()$ is an activation function; $y$ is the output data of the neuron, see Eq. (5).

$$S = \sum_{i=1}^{n}(w_i x_i) + b_1 , \qquad (4)$$

$$y = h(S) . \qquad (5)$$

Activation functions employed in neural networks are mandated to possess nonlinearity; otherwise, the purpose of multilayer hidden layers would be compromised [54]. Commonly employed activation functions encompass Sigmoid, Tanh, Relu, and Elu. The equations for these activation functions can be expressed as Equations (6) ~ (9). Moreover, visual representations of the Sigmoid, Tanh, Relu, and Elu functions are provided in Figure 2.

Sigmoid:
$$a(x) = \frac{1}{1+e^{-x}}, \tag{6}$$

Tanh:
$$a(x) = \frac{e^x - e^{-x}}{e^x + e^{-x}}, \tag{7}$$

Relu:
$$a(x) = \max(x, 0), \tag{8}$$

Elu:
$$a(x) = \begin{cases} e^x - 1 & x < 0 \\ x & x \geq 0 \end{cases}, \tag{9}$$

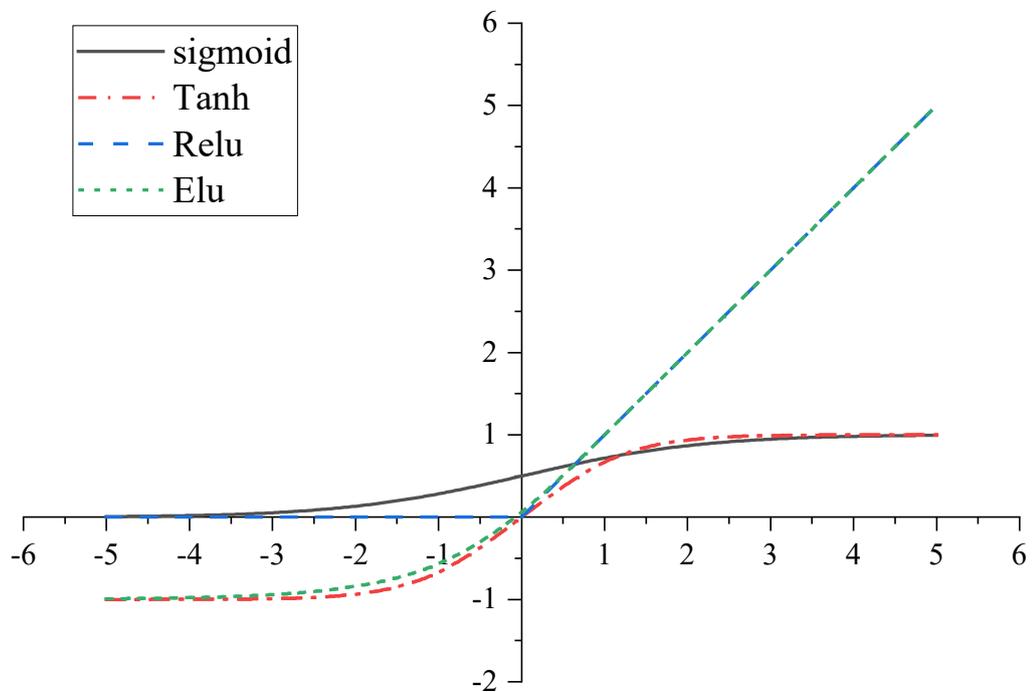

Figure 2 Sigmoid, Tanh, Relu and Elu functions

Figure 2 illustrates a neural network architecture, which typically comprises the input, hidden, and output layers. Neural networks are underpinned by the universal approximation theorem [55, 56], establishing their utility in predictive modeling, adaptive control, and scenarios where training with datasets is feasible. A neural network can infer insights from intricate and seemingly disparate datasets. Conceptually, a neural network can be envisioned as a trainable function. With an adequate volume of training data and an appropriate number of neurons, a neural network can acquire the capacity to learn and represent complex functions.

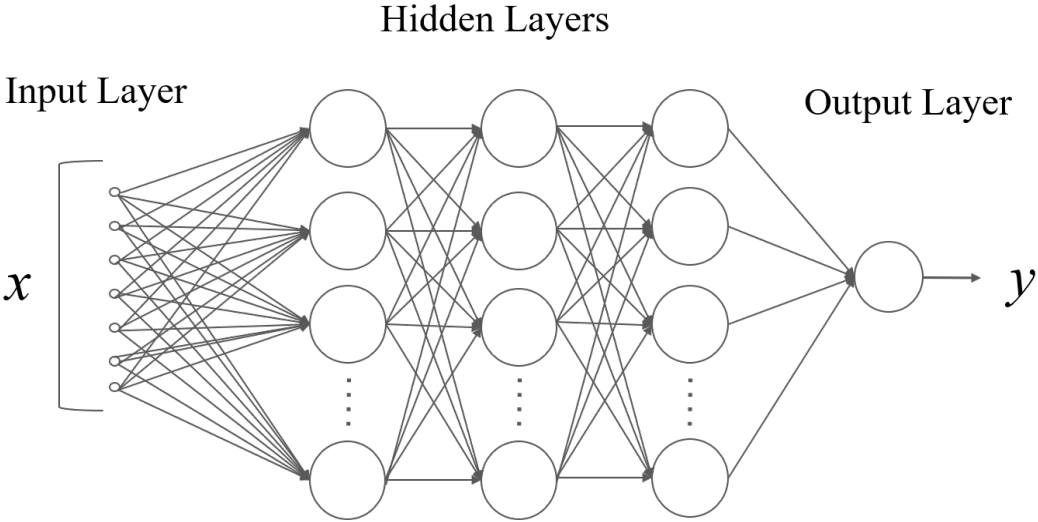

Figure 3. Neural network architectures.

2.3 Stochastic gradient descent

The training process of machine learning can be viewed as a problem of finding parameters $f(x;\theta)$ in the optimal model $\theta$, which optimization algorithms can learn. Gradient descent is the most commonly used optimization algorithm in machine learning. The optimization idea of the gradient descent method is to use the negative gradient direction of the current position as the search direction because that direction is the fastest descent direction of the current situation. The closer the gradient descent method is to the target value, the smaller the step size is. Several improved gradient descent methods have been developed based on gradient descent [57], such as batch gradient descent (BGD) [58], and stochastic gradient descent (SGD) [59, 60]. Regarding the optimization objective for each iteration, the batch gradient descent method is an average loss function over all samples, and the stochastic gradient descent method is a loss function over a single piece. The stochastic gradient descent method is widely used because it is relatively simple to implement and converges quickly in the early stages of computation. Compared with the batch gradient descent method, the stochastic gradient descent method incorporates random noise. This stochasticity often leads to faster convergence and the ability to escape local minima more effectively, making it suitable for a wide range of optimization problems. The algorithm of the SGD is shown in Program 1.

| **Program 1** Algorithm of stochastic gradient descent method |
|---|
| Input training set $\mathcal{D} = \left\{\left(\boldsymbol{x}^{(n)}, y^{(n)}\right)\right\}_{n=1}^{N}$, validation set $\mathcal{V}$, learning rate $\alpha$ |
| 1 Random initialization of parameter $\theta$; <br> 2 **repeat** <br> 3    Random sorting of the sample data within the training set $\mathcal{D}$; <br> 4    **for** $n = 1 \cdots N$ do <br> 5      Select the sample data $(x^{(n)}, y^{(n)})$ from the training set $\mathcal{D}$; <br> 6      $\theta \leftarrow \theta - \alpha \dfrac{\partial \mathcal{L}\left(\theta; x^{(n)}, y^{(n)}\right)}{\partial \theta}$, and update the parameter $\theta$; <br> 7    **end** <br> 8 **until** The error rate of model $f(x;\theta)$ on the validation dataset $\mathcal{V}$ no longer decreases; |
| Output parameter $\theta$ |

2.4 Automatic differentiation

    Automatic differentiation (AD) [61] is a computational technique designed to enable the precise and efficient computation of function derivatives through computer code [62]. Operating as an intermediary between symbolic and numerical differentiation, AD represents a well-balanced approach to derivative computation. At its core, AD entails the decomposition of differentials, as dictated by the chain rule. AD furnishes numerical derivatives by leveraging the symbolic differentiation rules [63]. It harnesses the inherent potency of the chain rule in calculus to break down differentials into elementary operations. AD achieves high accuracy in derivative calculations through the iterative application of the chain rule.

    The central concept revolves around creating a computational graph representing the function, with nodes symbolizing mathematical operations and edges denoting dependencies between variables. The AD process comprises two essential phases: forward propagation and backward propagation. During forward

propagation, input values traverse the computational graph from left to right, following solid lines, as illustrated in Figure 4. The corresponding mathematical operation is executed at each node, and intermediate results are stored for subsequent use in the backward propagation phase. In backward propagation, also known as backpropagation, derivatives are computed reversely, moving from right to left along dashed lines, as depicted in Figure 4.

The chain rule is systematically applied to ascertain the local derivatives of each operation about its inputs. These local derivatives are subsequently multiplied together to yield the derivative of the output concerning the input variables. AD computes differentiations with one forward and one backward propagation, regardless of the input's dimensionality [64]. The critical advantage of AD lies in its capability to calculate derivatives with a computational complexity that scales linearly with the number of operations in the function rather than being contingent on the input's dimensionality. This characteristic renders AD particularly advantageous in scenarios involving high-dimensional input spaces, where conventional methods like symbolic differentiation or numerical differentiation grapple with substantial challenges attributed to the inherent "dimensionality curse."

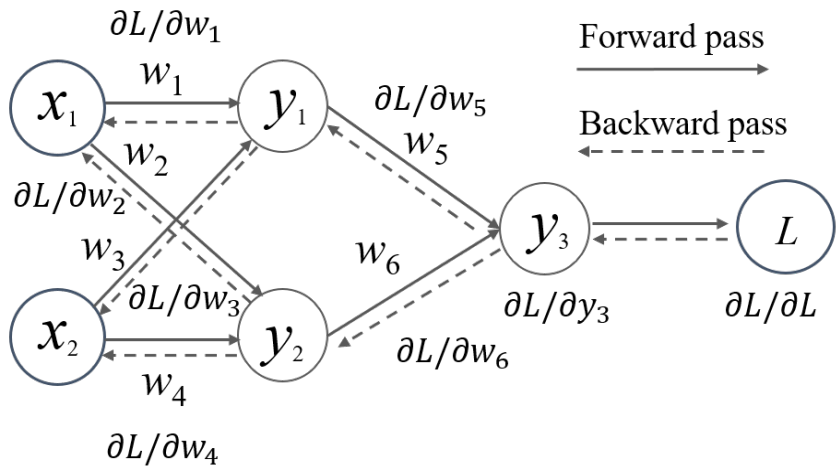

Figure 4. Overview of backpropagation.

## 3  Physics-informed neural network for seepage problems

The governing equation encapsulates the universal law adhered to by the head function, as represented in Equation (1). The precise form of the water head function can be derived by incorporating temporal and spatial initial and boundary conditions following the specific problem at hand. In the context of steady-state seepage, the head distribution remains unaltered to time, necessitating the imposition of only boundary conditions. However, for transient seepage problems, it becomes imperative to establish both the boundary conditions and the initial conditions.

The initial condition is the water head distribution in the seepage domain at the initial moment, and it can be expressed as

$$h\big|_{t=0} = h(x, y, 0). \tag{10}$$

The boundary condition describes the conditions of the water head $h$ on the boundary of the seepage region, indicating the interaction between the water head in the seepage region and the external system. There are three boundary conditions: the Dirichlet condition, the Neumann condition, and the Cauchy condition. For the time-dependent problem, we consider the time $t$ as a particular component of $x$ so that the initial condition can be used as a specific type of Dirichlet boundary condition.

The goal is to approximate $h$ with an approximating function $f(t, x; \theta)$ given by a deep neural network with parameter set $\theta$. The architecture of PINN for the seepage analysis, as shown in Figure 5. The loss functional for the associated training problem consists of three parts:

A measure of how well the approximation satisfies the differential equation

$$Loss_f = \left\|(\partial_t - \Delta) f(t, x; \theta)\right\|^2_{[0,T] \times \Omega, \nu_1}. \tag{11}$$

A measure of how well the approximation satisfies the boundary condition

$$Loss_{BC} = \|f(t, x; \theta) - g(t, x)\|^2_{[0,T] \times \partial\Omega, \nu_2}. \tag{12}$$

A measure of how well the approximation satisfies the initial condition

$$Loss_{IC} = \|f(0, x; \theta) - h_0(x)\|^2_{\Omega, \nu_3}. \tag{13}$$

In all three terms above the error is evaluated in terms of $L^2$-norm. Combining the three terms above gives us the cost functional associated with training the neural network:

$$L(\theta) = Loss_f + Loss_{BC} + Loss_{IC}. \tag{14}$$

The next step is to minimize the loss functional using the SGD. The overview of the whole algorithm is shown in Program 1.

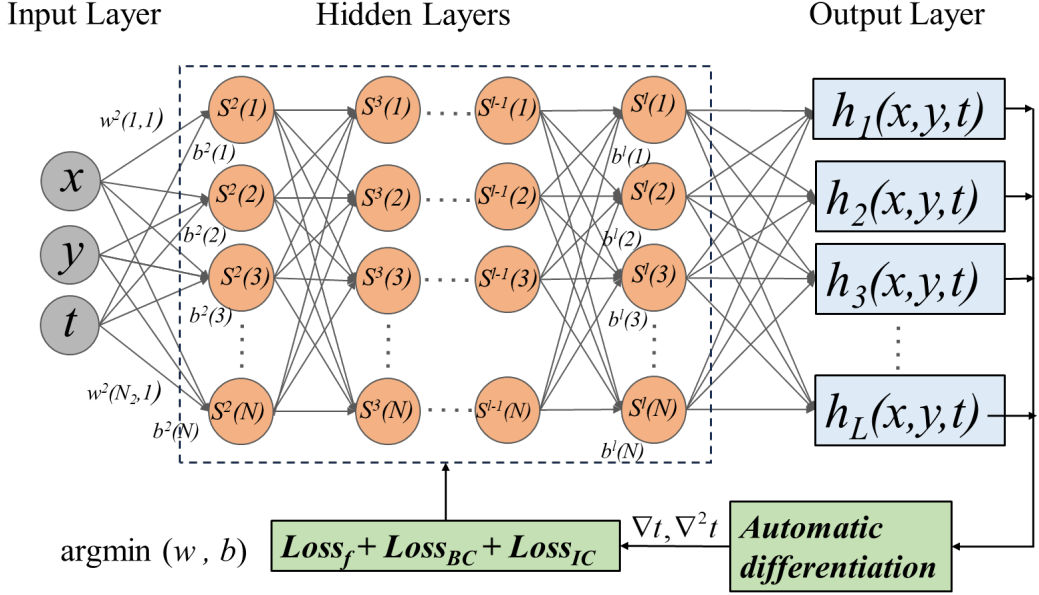

Figure 5 Architecture of PINN for the seepage problems.

## 4 Numerical tests

In this work, we explored steady-state, transient, and free-surface seepage problems through the solution of four numerical examples. To assess the accuracy of the PINN, we conducted a comparative analysis with the FEM, which was implemented using the commercial software ABAQUS. To validate our findings, we evaluated the relative error in the $L_2$ norm of the hydraulic head as follows:

$$e_{L^2} = \|h_{num} - h_{ref}\|_{L^2(\Omega)} = \frac{\sqrt{\int_\Omega (h_{num} - h_{ref})^T (h_{num} - h_{ref}) d\Omega}}{\sqrt{\int_\Omega h_{ref}^T h_{ref} d\Omega}}, \quad (15)$$

where $h_{num}$ denotes the numerical solution and $h_{ref}$ denotes the reference solution.

Table 1. Architecture of the PINN for different numerical tests

| Numerical tests | Hidden layers | Neurons on each layers | Optimizer | Learning rate | Iterations |
|---|---|---|---|---|---|
| Test 1 | 4 | 50 | Adam, L-BFGS | 0.001 | 20000 |
| Test 2 | 4 | 50 | Adam, L-BFGS | 0.001 | 20000 |
| Test 3 | 8 | 50 | Adam, L-BFGS-B | 0.001 | 20000 |
| Test 4 | 8 | 50 | Adam, L-BFGS-B | 0.001 | 20000 |

4.1 Numerical test 1: Steady-state seepage problem for rectangular plate with holes

In the first numerical test, we compute the steady-state seepage problem for a rectangular plate with holes. As shown in Figure 6, the side length of a square rectangular plate with holes is 2 m, and the hole's radius is 0.5 m. Four monitoring points, a (-0.75, 0), b (0, 0.75), c (0.75, 0), and d (0, -0.75) are arranged on the board. The water head at the boundary AB is 0.4 m, the water head at the boundary CD is 0.3 m, and the water head at the boundary of the circular excavation hole is 0.8 m. The boundaries AD and BC are impermeable boundaries. The boundary conditions are shown in Equations (16) ~ (18). The permeability coefficient is $k = 0.001 m/s$.

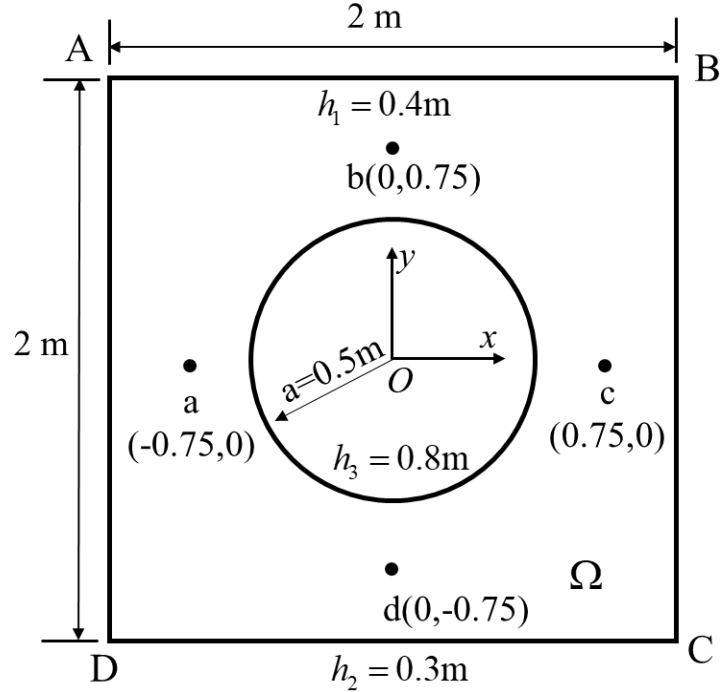

Figure 6. Geometry and boundary conditions of a rectangular plate with holes.

The partial differential equations of steady-state seepage problem can be written as

$$k(\frac{\partial^2 h}{\partial x^2} + \frac{\partial^2 h}{\partial y^2}) = 0, \quad x, y \in \Omega, \tag{2}$$

the boundary conditions:

$$h_1(x_1, 1) = 0.4m, \quad x_1 \in (-1, 1), \tag{16}$$

$$h_2(x_2, -1) = 0.3m, \quad x_2 \in (-1, 1), \tag{17}$$

$$h_3(x_3, y_3) = 0.8m, \quad \text{on } (x_3)^2 + (y_3)^2 = 0.5^2. \tag{18}$$

To verify the accuracy of the PINN, we need to get a reliable reference solution using the FEM. The model of the FEM is shown in Figure 7, which consists of 2128 nodes and 2016 CPE4P elements. Moreover, we solve the steady-state seepage problem using the PINN. The neural network we use consists of an input layer consisting of two neurons, four hidden layers with 50 neurons each, and an output layer. There are 1200 training points evenly distributed on the geometric region, of which 200 training points are distributed on the boundary, and 1000 training points are distributed on the region. Now, we have the PDE problem and the network. We build a Model and then define the optimizer and learning rate. We train the model using Adam and a learning rate of 0.001 for 20,000 iterations, and then, we train the network using L-BFGS to achieve even more minor losses.

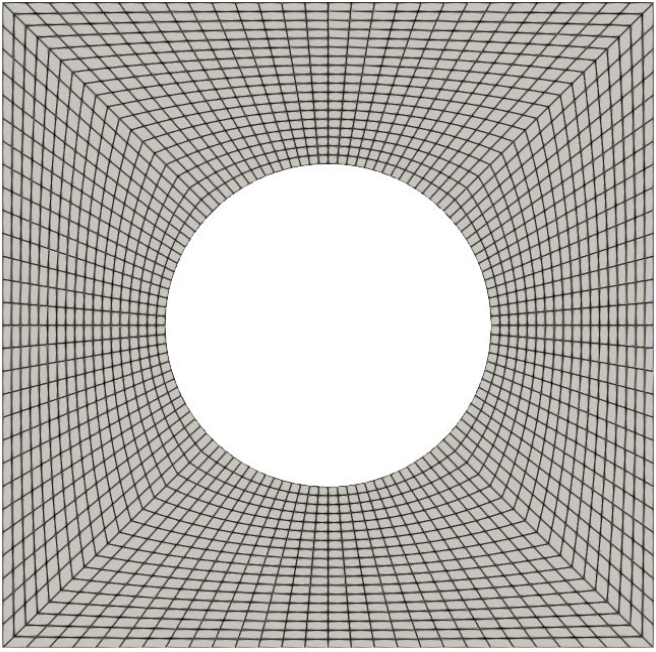

Figure 7. Model of the FEM for rectangular plate with holes.

Figure 8 illustrates the hydraulic head contours for both the PINN and the FEM. This visualization highlights a remarkable similarity in the outcomes obtained through these two approaches. Moreover, Figure 8 (c) shows the relative error between the PINN and the FEM. The relative error is predominantly found at the interior hollow circular boundary, where the maximum relative error is $9 \times 10^{-4}$. Therefore, there is minimal difference between the solution results of the PINN and the FEM, indicating that they can be readily applied to complex boundary conditions. Table 2 compares the PINN and the FEM results for four monitor points. The comparison between the solution results obtained via the PINN and those derived from the FEM at the four observation points reveals a remarkable similarity. The relative error between the FEM and the PINN results is a mere 0.0254%.

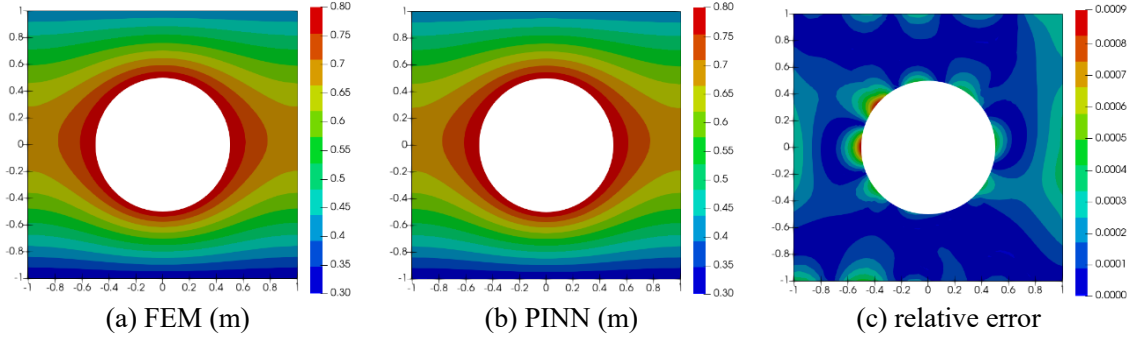

(a) FEM (m)　　　　　　　　(b) PINN (m)　　　　　　　　(c) relative error

Figure 8. Hydraulic head distribution and relative errors of the steady-state seepage problem with holes in a rectangular plate; (a) the FEM. (b) the PINN. (c) relative error of between the FEM and the PINN.

Table 2.

| Method | Monitor point | | | | Relative error $e_{L^2}$ (%) |
|---|---|---|---|---|---|
| | a | b | c | d | |
| FEM | 0.714003 m | 0.581901 m | 0.714003 m | 0.527604 m | - |
| PINN | 0.714316 m | 0.581933 m | 0.713927 m | 0.527577 m | 0.0254 |

4.2 Numerical test 2: steady-state seepage problem of concrete dam foundation

This test considers a standard concrete dam foundation steady-state saturated seepage problem. The geometric model and boundary conditions are shown in Figure 9. We assume that the dam is impervious to water. In order to verify the accuracy of the proposed method, three monitoring points, a(100,80), b(120,80), and c(140,80), are arranged on the boundary EF, as shown in Figure 9. The boundaries BC, AE, EF, and DF are defined as impermeable boundaries. The hydraulic head of AB is 80 m, and the hydraulic head of CD is 20 m. The coefficient of permeability is $k_x = k_y = 1\times10^{-6}\,m/s$. The partial differential equations and boundary conditions can be written as

$$\Delta h(x, y) = 0, \quad x, y \in \Omega, \qquad (3)$$

$$h(x_1, 80) = 80\mathrm{m}, \quad x_1 \in [0, 80], \qquad (20)$$

$$h(x_2, 80) = 20\mathrm{m}, \quad x_2 \in [160, 240]. \qquad (21)$$

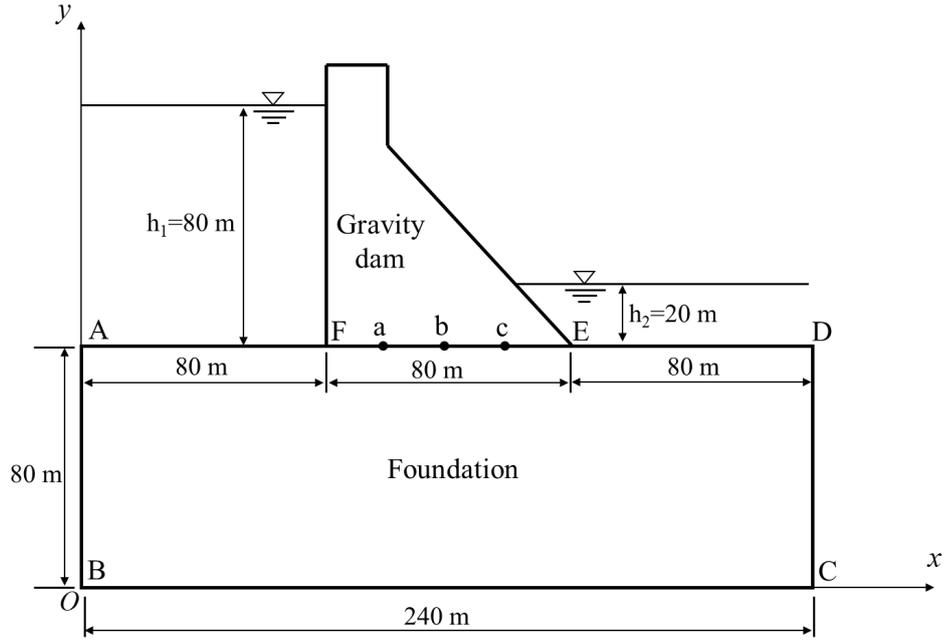

Figure 9. Geometry and boundary conditions of concrete dam foundation.

In this numerical test, a six-layer neural network is used, where the first layer is an input layer with two neurons, the second through fifth layers are hidden layers with 50 neurons, and the sixth layer is an output layer with one neuron. The geometric domain of the gravity dam foundation is uniformly distributed with 2600 training points, of which 600 training points are on the boundary, and 2000 training points are within the domain. We first use the Adam optimizer at a learning rate of 0.005 for 20,000 iterations as the first stage of training, and then, to achieve even more minor losses, we use the L-BFGS optimizer for the second stage of training.

The hydraulic distribution of the PINN is shown in Figure 10 (b), the relative errors between the FEM results and PINN results are shown in Figure 10 (c). It is evident from Figure 10 (c) that the head distribution obtained from PINN aligns closely with the FEM results, and the maximum head error is a mere 0.217. More significant errors occur mainly in the parts where the head boundary conditions change drastically, such as points F and E. Comparing the PINN and the FEM of the three monitoring points with the analytical solution, as shown in Table 3. The relative errors of the FEM and the PINN are 3.82% and 1.38%, respectively. Hence, the PINN shows a greater accuracy than the FEM.

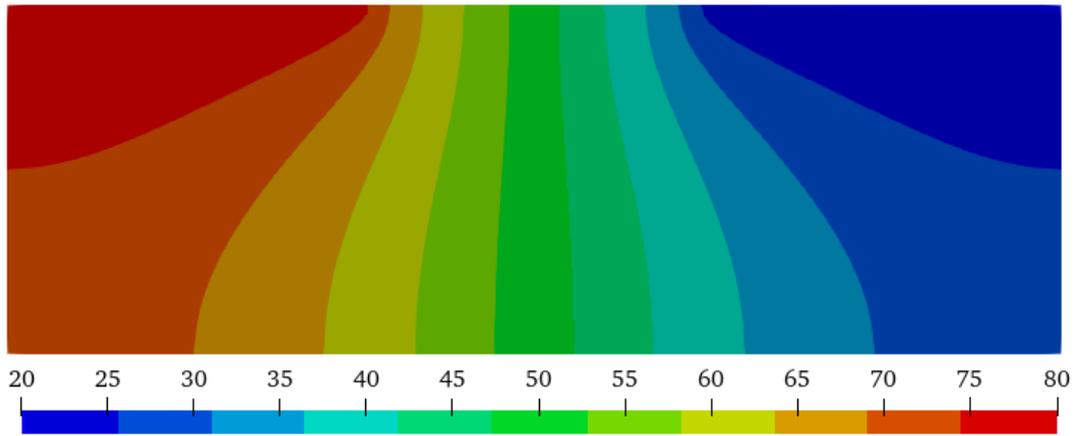

(a) FEM (m)

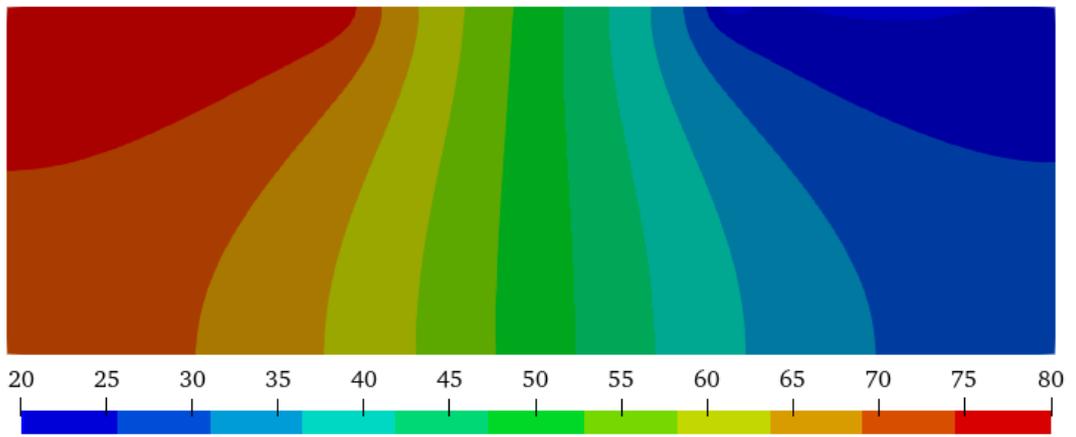

(b) PINN (m)

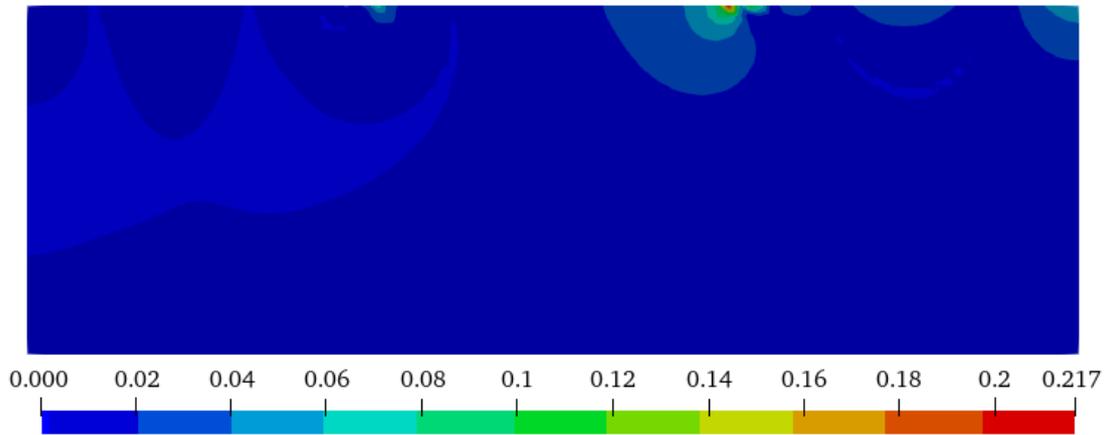

(c) Relative error

Figure 10. Results and relative errors for a steady-state seepage problem in the gravity dam foundation: (a) the FEM . (b) the PINN. (c) Relative error

Table 3. Comparison of water head by different methods

| Method | Monitor point | | | Relative error $e_{L^2}$ (%) |
| --- | --- | --- | --- | --- |
| | 1 | 2 | 3 | |
| Analytical solution (m) | 60 | 50 | 40 | - |
| FEM (m) [65] | 62.45 | 49.90 | 37.39 | 3.82 |
| PINN (m) | 60.6224 | 50.8095 | 40.6531 | 1.38 |

4.3 Numerical test 3: free-surface seepage problems

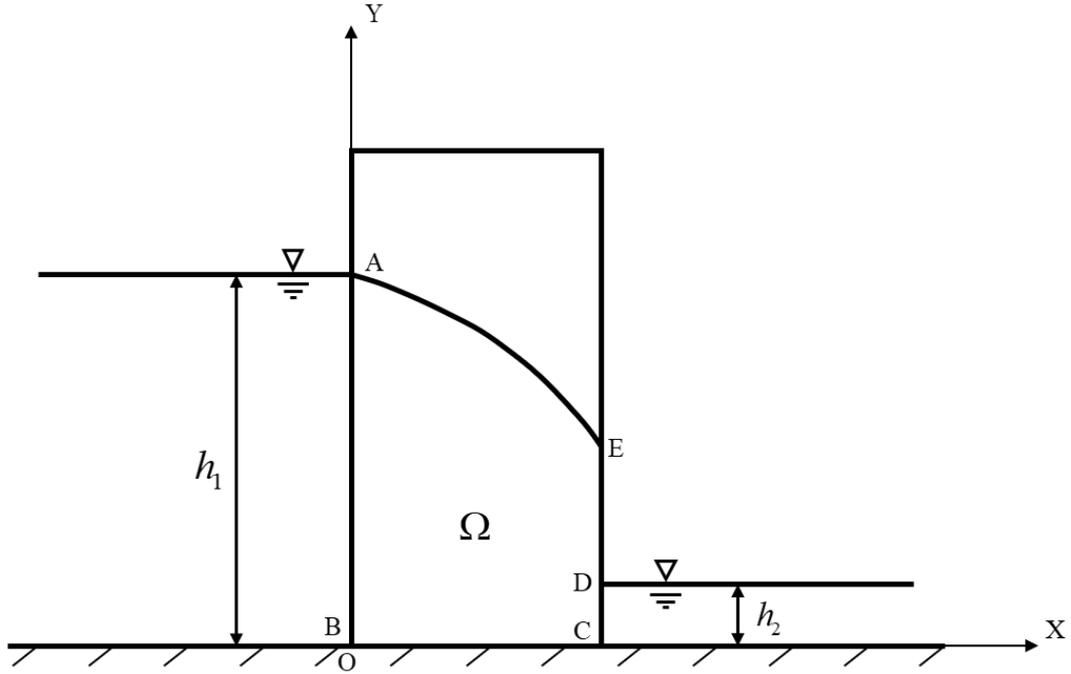

Figure 11. Free-surface seepage problem.

To validate the feasibility of PINN and assess the accuracy of solving the free-surface, we have chosen a representative example supported by experimental data. The geometry model and boundary conditions are illustrated in Figure 11. The boundary conditions include AB as the upstream head boundary condition, CD as the downstream head boundary condition with a constant value, BC representing the impermeable bottom boundary, AE signifying the seepage free-surface, and DE representing the seepage overflow surface, and E as the overflow point. The governing equation and boundary conditions are written as

$$h_1 = 6\text{m} \quad \text{on AB}, \tag{4}$$

$$h_2 = 1\text{m} \quad \text{on CD}, \tag{5}$$

$$\frac{\partial h}{\partial n} = 0 \quad \text{on BC}, \tag{6}$$

$$h_3 = y \quad \text{on DE}, \tag{7}$$

$$\frac{\partial h}{\partial n} = 0 \text{ and } h_4 = y \quad \text{on AE}. \tag{8}$$

In this numerical experiment, we employed a 10-layer neural network architecture comprising an initial input layer, followed by eight hidden layers (the second to ninth layers), culminating in the tenth layer serving as the output layer. The neural network produced outputs corresponding to the partial differential equations and the boundary conditions associated with the free-surface. Regarding the neural network training, we initially conducted 20,000 iterations using the Adam optimizer with a learning rate 0.001. Subsequently, we continued the training process by employing the L-BFGS-B optimizer to minimize the loss function further. The trained model was then utilized to make predictions and compared against experimental data. The results of this comparison are presented in Table 4 and Figure 12. For the problem of seepage with a free-surface, the solutions generated by the PINN closely align with the experimental observations. Moreover, the average relative error of the PINN and the FEM is 1.24%, and 2.53%, respectively. The accuracy of the PINN hence is greater than the FEM. This demonstrates the efficacy of PINN in effectively addressing complex boundary conditions encountered in free-surface seepage problems.

Table 4. Comparison of water head by different methods

| Location (x/m) | Experiment (x/m)[66] | PINN (x/m) | Relative error between PINN and experiment (%) | FEM(x/m)[66] | Relative error between FEM and experiment (%) |
|---|---|---|---|---|---|
| 0.0 | 6 | 6.02 | 0.33 | | |
| 0.5 | 5.82 | 5.81 | 0.17 | | |
| 1.0 | 5.63 | 5.57 | 1.07 | 5.67 | 0.71 |
| 1.5 | 5.37 | 5.31 | 1.12 | | |
| 2.0 | 5.10 | 5.07 | 0.59 | 5.24 | 2.75 |
| 2.5 | 4.74 | 4.79 | 1.05 | | |
| 3.0 | 4.38 | 4.46 | 1.83 | 4.47 | 2.05 |
| 3.5 | 3.82 | 3.87 | 1.31 | | |
| 4.0 | 3.25 | 3.13 | 3.69 | 3.40 | 4.62 |

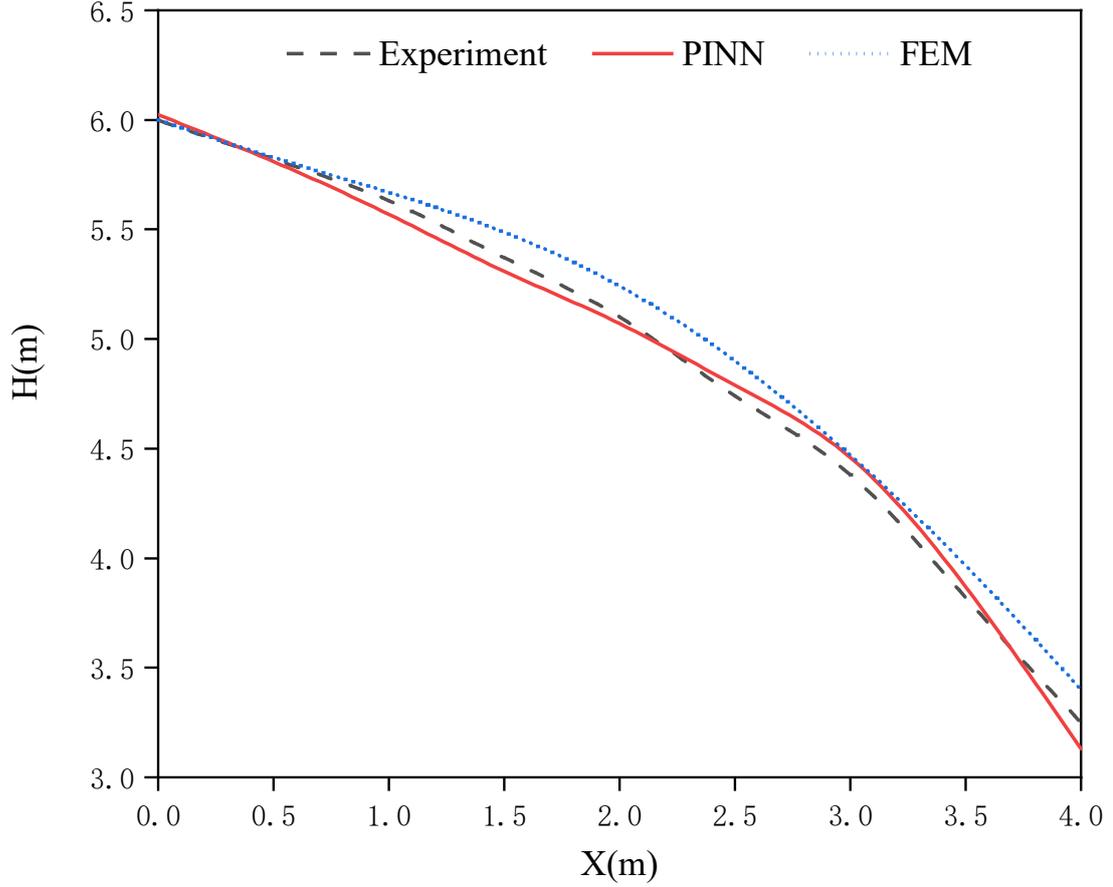

Figure 12. Comparison of different methods for solving the free-surface seepage problem.

4.4 Numerical test 4: transient seepage problem

The last test considers a transient seepage problem in a two-dimensional homogeneous porous medium. As shown in Figure 13, the flow region is a 2 × 2 square area with a permeability coefficient of $k = 0.1 m/s$. The governing equation, initial conditions, and boundary conditions are as follows:

the partial differential equations:

$$k(\frac{\partial^2 h}{\partial x^2} + \frac{\partial^2 h}{\partial y^2}) = \frac{\partial h}{\partial t}, \tag{9}$$

the boundary conditions:

$$h(0, y, t) = 0 \text{m}, \tag{10}$$

$$h(2, y, t) = 3 \text{m}, \tag{11}$$

the initial conditions:

$$h(x, y, t) = 0 \text{m} \quad x \in [0, 1), \tag{12}$$

$$h(x, y, t) = 3(x - 1) \quad x \in [1, 2], \tag{13}$$

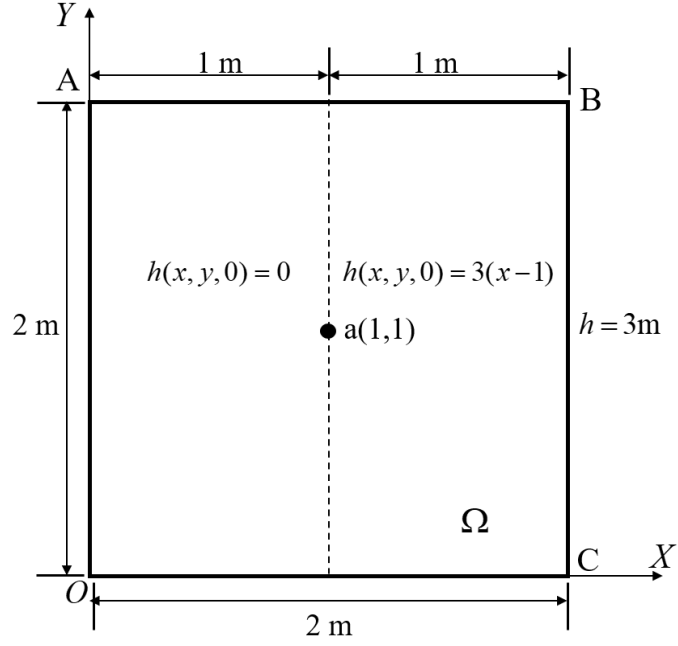

Figure 13. transient saturated seepage problem

In this numerical test, we employed a 10-layer neural network architecture. The initial layer comprises three input neurons, followed by seven hidden layers (from the second to eighth layers), each consisting of 50 neurons. The tenth and final layer serves as the output layer with one neuron. Two thousand training points were distributed within the computational domain, with an additional 500 nodes situated along the boundary. This particular problem involves initial conditions, and the initial training points were organized accordingly. Similar to our previous numerical test, we utilized the quadratic training methodology. Specifically, we initiated training with the Adam optimizer, conducting 20,000 iterations while maintaining a learning rate 0.001. Subsequently, we further refined the training process using the L-BFGS-B optimizer to achieve even lower loss values.

We also conducted a comparative analysis by solving this problem using the finite element method (FEM) alongside the Physics-Informed Neural Network (PINN) approach. The FEM model employed a total of 1600 CPE4P elements and 1681 nodes. Figure 14 illustrates the results obtained from PINN and FEM at various time intervals and the corresponding error between the two methods. Remarkably, the results from both approaches exhibit a high degree of consistency and alignment. Furthermore, Figure 15 depicts the temporal evolution of the head at monitoring point 'a,' employing both the PINN and FEM. Notably, the solutions generated by these two methods show exceptional correspondence and agreement.

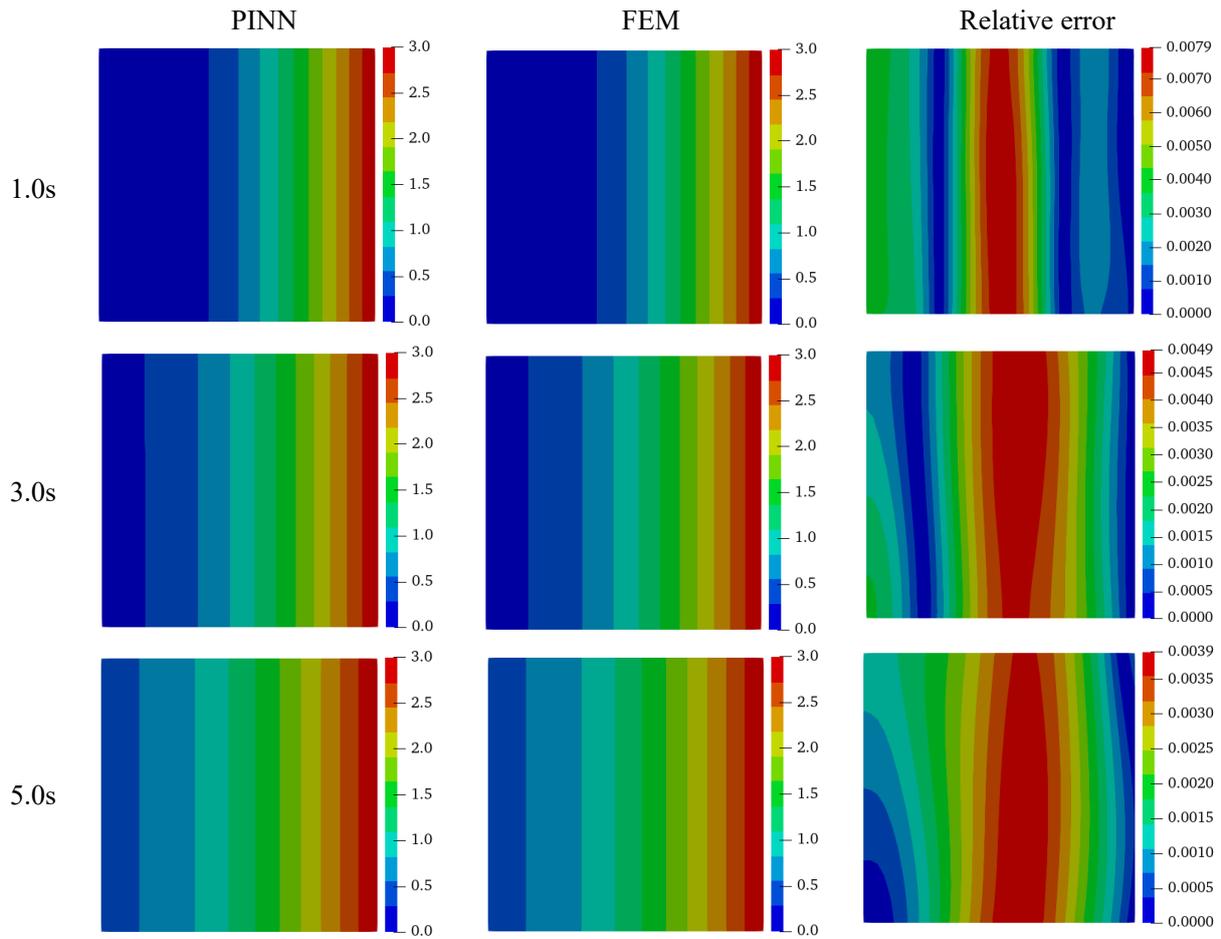

Figure 14. Results of the PINN, calculation results of the FEM, relative error between calculation results of different methods

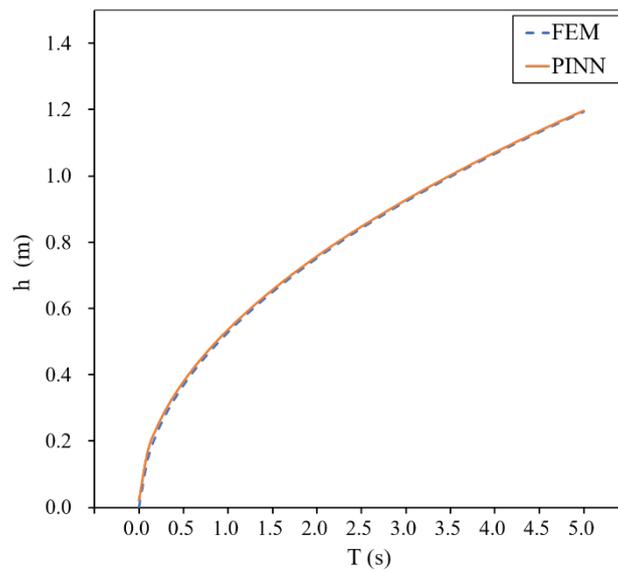

Figure 15 Comparison of solution results of monitoring point a using the PINN and the FEM.

# 5  Conclusions

This study advocates the utilization of the PINN for the resolution of seepage problems. A series of benchmark problems are tackled employing the PINN and the FEM, affirming the accuracy and viability of the proposed computational methodology.

The work reveals a remarkable similarity between the solutions obtained through the PINN and the FEM, with the former exhibiting superior accuracy in steady-state problems compared to the FEM. Transient problems exhibit excellent agreement between the PINN and the FEM results. Notably, PINN demonstrates exceptional efficacy in handling intricate boundary conditions inherent in free-surface seepage problems. The accuracy of PINN surpasses that of FEM in addressing free-surface seepage problems. Consequently, the presented approach showcases robustness and precision in effectively addressing a spectrum of seepage challenges.

**Acknowledgement**

Project on Excellent Post-graduate Dissertation of Hohai University (422003509), the Xing Dian Talent Support Program of Yunnan Province (XDYC-QNRC-2022-0764) and the Research on Key Technologies of Green Energy Security and Sustainable Development in Yunnan Province (2021KXJGZS01) support this study.

**Reference**

1. Huangfu M, Wang M-S, Tan Z-S, Wang X-Y (2010) Analytical solutions for steady seepage into an underwater circular tunnel. Tunnelling and Underground Space Technology 25:391–396. https://doi.org/10.1016/j.tust.2010.02.002

2. Rice JD, Duncan JM (2010) Deformation and Cracking of Seepage Barriers in Dams due to Changes in the Pore Pressure Regime. Journal of Geotechnical and Geoenvironmental Engineering 136:16–25. https://doi.org/10.1061/(ASCE)GT.1943-5606.0000241

3. Jeyisanker K, Gunaratne M (2009) Analysis of water seepage in a pavement system using the particulate approach. Computers and Geotechnics 36:641–654. https://doi.org/10.1016/j.compgeo.2008.09.002

4. Ma L, Huang C, Liu Z-S, Morin KA, Aziz M, Meints C (2020) Artificial Neural Network for Prediction of Full-Scale Seepage Flow Rate at the Equity Silver Mine. Water Air Soil Pollut 231:179. https://doi.org/10.1007/s11270-020-04541-x

5. Li GC, Desai CS (1983) Stress and Seepage Analysis of Earth Dams. Journal of Geotechnical Engineering 109:946–960. https://doi.org/10.1061/(ASCE)0733-9410(1983)109:7(946)

6. Yang C, Sheng D, Carter JP (2012) Effect of hydraulic hysteresis on seepage analysis for unsaturated

soils. Computers and Geotechnics 41:36–56. https://doi.org/10.1016/j.compgeo.2011.11.006

7. Fahimifar A, Ghadami H, Ahmadvand M (2015) An elasto-plastic model for underwater tunnels considering seepage body forces and strain-softening behaviour. European Journal of Environmental and Civil Engineering 19:129–151. https://doi.org/10.1080/19648189.2014.939305

8. Barua G, Tiwari KN (1995) Analytical Solutions of Seepage into Ditches from Ponded Fields. Journal of Irrigation and Drainage Engineering 121:396–404. https://doi.org/10.1061/(ASCE)0733-9437(1995)121:6(396)

9. Li X, Du S, Chen B (2017) Unified analytical solution for deep circular tunnel with consideration of seepage pressure, grouting and lining. J Cent South Univ 24:1483–1493. https://doi.org/10.1007/s11771-017-3552-3

10. Xin P, Dan H-C, Zhou T, Lu C, Kong J, Li L (2016) An analytical solution for predicting the transient seepage from a subsurface drainage system. Advances in Water Resources 91:1–10. https://doi.org/10.1016/j.advwatres.2016.03.006

11. Damlamian A (1984) Weak and Variational Methods for Moving Boundary Problems (C. M. Elliott and J. R. Ockendon). SIAM Rev 26:137–138. https://doi.org/10.1137/1026023

12. Al-Janabi AMS, Ghazali AH, Ghazaw YM, Afan HA, Al-Ansari N, Yaseen ZM (2020) Experimental and Numerical Analysis for Earth-Fill Dam Seepage. Sustainability 12:2490. https://doi.org/10.3390/su12062490

13. Akay O, Özer AT, Fox GA, Bartlett SF, Arellano D (2013) Behavior of sandy slopes remediated by EPS-block geofoam under seepage flow. Geotextiles and Geomembranes 37:81–98. https://doi.org/10.1016/j.geotexmem.2013.02.005

14. Smith AJ, Herne DE, Turner JV (2009) Wave effects on submarine groundwater seepage measurement. Advances in Water Resources 32:820–833. https://doi.org/10.1016/j.advwatres.2009.02.003

15. Isiorho SA, Meyer JH (1999) The Effects of Bag Type and Meter Size on Seepage Meter Measurements. Groundwater 37:411–413. https://doi.org/10.1111/j.1745-6584.1999.tb01119.x

16. Sheng H, Yang C (2021) PFNN: A penalty-free neural network method for solving a class of second-order boundary-value problems on complex geometries. Journal of Computational Physics 428:110085. https://doi.org/10.1016/j.jcp.2020.110085

17. Kim J, Kim D, Choi H (2001) An Immersed-Boundary Finite-Volume Method for Simulations of Flow in Complex Geometries. Journal of Computational Physics 171:132–150. https://doi.org/10.1006/jcph.2001.6778

18. Darbandi M, Torabi SO, Saadat M, Daghighi Y, Jarrahbashi D (2007) A moving-mesh finite-volume method to solve free-surface seepage problem in arbitrary geometries. International Journal for Numerical and Analytical Methods in Geomechanics 31:1609–1629. https://doi.org/10.1002/nag.611

19. Ying GUO, Xin-liang J, Dong-bo C a. O, Tie-jun B a. I, Guang-yi ZHU, Chun F (2018) A FINITE VOLUME NUMERICAL SIMULATION METHOD FOR ROCK MASS STRENGTH WEAKENING BY SEEPAGE WATER ABSORBING. gclx 35:139–149. https://doi.org/10.6052/j.issn.1000-4750.2017.03.0209

20. Volz C, Frank P-J, Vetsch DF, Hager WH, Boes RM (2017) Numerical embankment breach modelling including seepage flow effects. Journal of Hydraulic Research 55:480–490. https://doi.org/10.1080/00221686.2016.1276104

21. Neuman SP, Witherspoon PA (1970) Finite Element Method of Analyzing Steady Seepage with a Free Surface. Water Resources Research 6:889–897. https://doi.org/10.1029/WR006i003p00889

22. Bathe K-J, Khoshgoftaar MR (1979) Finite element free surface seepage analysis without mesh iteration. International Journal for Numerical and Analytical Methods in Geomechanics 3:13–22. https://doi.org/10.1002/nag.1610030103

23. Fukuchi T (2016) Numerical analyses of steady-state seepage problems using the interpolation finite difference method. Soils and Foundations 56:608–626. https://doi.org/10.1016/j.sandf.2016.07.003

24. Rohaninejad M, Zarghami M (2012) Combining Monte Carlo and finite difference methods for effective simulation of dam behavior. Advances in Engineering Software 45:197–202. https://doi.org/10.1016/j.advengsoft.2011.09.023

25. Caffrey J, Bruch JC (1979) Three-dimensional seepage through a homogeneous dam. Advances in Water Resources 2:167–176. https://doi.org/10.1016/0309-1708(79)90032-0

26. Liu Z-F, Wang X-H (2013) Finite analytic numerical method for two-dimensional fluid flow in heterogeneous porous media. Journal of Computational Physics 235:286–301. https://doi.org/10.1016/j.jcp.2012.11.001

27. Wang H, Qin Q-H, Kang Y-L (2006) A meshless model for transient heat conduction in functionally graded materials. Comput Mech 38:51–60. https://doi.org/10.1007/s00466-005-0720-3

28. XueHong W, ShengPing S, WenQuan T (2007) Meshless Local Petrov-Galerkin Collocation Method for Two-dimensional Heat Conduction Problems. 12

29. Xiao J-E, Ku C-Y, Huang W-P, Su Y, Tsai Y-H (2018) A Novel Hybrid Boundary-Type Meshless Method for Solving Heat Conduction Problems in Layered Materials. Applied Sciences 8:1887. https://doi.org/10.3390/app8101887

30. Hackbusch W, Nowak ZP (1989) On the fast matrix multiplication in the boundary element method by panel clustering. Numer Math 54:463–491. https://doi.org/10.1007/BF01396324

31. AL-Jawary MA, Wrobel LC (2012) Radial integration boundary integral and integro-differential equation methods for two-dimensional heat conduction problems with variable coefficients. Engineering Analysis with Boundary Elements 36:685–695. https://doi.org/10.1016/j.enganabound.2011.11.019

32. Wang C, Khoo BC (2004) An indirect boundary element method for three-dimensional explosion bubbles. Journal of Computational Physics 194:451–480. https://doi.org/10.1016/j.jcp.2003.09.011

33. Yang Y, Zhang Z, Feng Y, Wang K (2022) A Novel Solution for Seepage Problems Implemented in the Abaqus UEL Based on the Polygonal Scaled Boundary Finite Element Method. Geofluids 2022:e5797014. https://doi.org/10.1155/2022/5797014

34. Lin S, Cao X, Zheng H, Li Y, Li W (2023) An improved meshless numerical manifold method for simulating complex boundary seepage problems. Computers and Geotechnics 155:105211. https://doi.org/10.1016/j.compgeo.2022.105211

35. Chaiyo K, Rattanadecho P, Chantasiriwan S (2011) The method of fundamental solutions for solving free boundary saturated seepage problem. International Communications in Heat and Mass Transfer 38:249–254. https://doi.org/10.1016/j.icheatmasstransfer.2010.11.022

36. Yang Y, Zhang Z, Feng Y, Yu Y, Wang K, Liang L (2021) A polygonal scaled boundary finite element method for solving heat conduction problems. arXiv preprint arXiv:210612283

37. Seifert C, Aamir A, Balagopalan A, Jain D, Sharma A, Grottel S, Gumhold S (2017) Visualizations of Deep Neural Networks in Computer Vision: A Survey. In: Cerquitelli T, Quercia D, Pasquale F (eds) Transparent Data Mining for Big and Small Data. Springer International Publishing, Cham, pp 123–144

38. Fathi E, Maleki Shoja B (2018) Chapter 9 - Deep Neural Networks for Natural Language Processing. In: Gudivada VN, Rao CR (eds) Handbook of Statistics. Elsevier, pp 229–316

39. Shin D, Kim H, Park K, Yi K (2020) Development of Deep Learning Based Human-Centered Threat Assessment for Application to Automated Driving Vehicle. Applied Sciences 10:253. https://doi.org/10.3390/app10010253

40. Bergen KJ, Johnson PA, de Hoop MV, Beroza GC (2019) Machine learning for data-driven discovery in solid Earth geoscience. Science 363:eaau0323. https://doi.org/10.1126/science.aau0323

41. Kong Q, Trugman DT, Ross ZE, Bianco MJ, Meade BJ, Gerstoft P (2018) Machine Learning in Seismology: Turning Data into Insights. Seismological Research Letters 90:3–14. https://doi.org/10.1785/0220180259

42. Brenner MP, Eldredge JD, Freund JB (2019) Perspective on machine learning for advancing fluid mechanics. Phys Rev Fluids 4:100501. https://doi.org/10.1103/PhysRevFluids.4.100501

43. Kochkov D, Smith JA, Alieva A, Wang Q, Brenner MP, Hoyer S (2021) Machine learning–accelerated computational fluid dynamics. Proceedings of the National Academy of Sciences 118:e2101784118. https://doi.org/10.1073/pnas.2101784118

44. Abadi M, Barham P, Chen J, Chen Z, Davis A, Dean J, Devin M, Ghemawat S, Irving G, Isard M, Kudlur M, Levenberg J, Monga R, Moore S, Murray DG, Steiner B, Tucker P, Vasudevan V, Warden P, Wicke M, Yu Y, Zheng X (2016) {TensorFlow}: A System for {Large-Scale} Machine Learning. pp 265–283

45. Paszke A, Gross S, Massa F, Lerer A, Bradbury J, Chanan G, Killeen T, Lin Z, Gimelshein N, Antiga L, Desmaison A, Kopf A, Yang E, DeVito Z, Raison M, Tejani A, Chilamkurthy S, Steiner B, Fang L, Bai J, Chintala S (2019) PyTorch: An Imperative Style, High-Performance Deep Learning Library. In: Advances in Neural Information Processing Systems. Curran Associates, Inc.

46. Haghighat E, Juanes R (2021) SciANN: A Keras/Tensorflow wrapper for scientific computations and physics-informed deep learning using artificial neural networks. Computer Methods in Applied Mechanics and Engineering 373:113552. https://doi.org/10.1016/j.cma.2020.113552

47. Cuomo S, Di Cola VS, Giampaolo F, Rozza G, Raissi M, Piccialli F (2022) Scientific Machine Learning Through Physics–Informed Neural Networks: Where we are and What's Next. J Sci Comput 92:88. https://doi.org/10.1007/s10915-022-01939-z

48. Yu J, Lu L, Meng X, Karniadakis GE (2022) Gradient-enhanced physics-informed neural networks for forward and inverse PDE problems. Computer Methods in Applied Mechanics and Engineering 393:114823. https://doi.org/10.1016/j.cma.2022.114823

49. Pu J, Song W, Wu J, Gou F, Yin X, Long Y (2022) PINN-Based Method for Predicting Flow Field Distribution of the Tight Reservoir after Fracturing. Geofluids 2022:e1781388. https://doi.org/10.1155/2022/1781388

50. Daolun L, Luhang S, Wenshu Z, Xuliang L, Jieqing T (2021) Physics-constrained deep learning for solving seepage equation. Journal of Petroleum Science and Engineering 206:109046. https://doi.org/10.1016/j.petrol.2021.109046

51. Gao Y, Qian L, Yao T, Mo Z, Zhang J, Zhang R, Liu E, Li Y (2023) An Improved Physics-Informed Neural Network Algorithm for Predicting the Phreatic Line of Seepage. Advances in Civil Engineering 2023:e5499645. https://doi.org/10.1155/2023/5499645

52. Kim S, Choi J-H, Kim NH (2022) Data-driven prognostics with low-fidelity physical information for digital twin: physics-informed neural network. Struct Multidisc Optim 65:255. https://doi.org/10.1007/s00158-022-03348-0

53. Bear J, Verruijt A (1987) Modeling Groundwater Flow and Pollution. Springer Science & Business Media

54. Rasamoelina AD, Adjailia F, Sincak P (2020) A Review of Activation Function for Artificial Neural Network. In: 2020 IEEE 18th World Symposium on Applied Machine Intelligence and Informatics (SAMI). IEEE, Herlany, Slovakia, pp 281–286

55. Cybenko G V (1993) Approximation by superpositions of a sigmoidal function. Mathematics of Control Signals and Systems 5:17–28

56. Hornik K (1991) Approximation capabilities of multilayer feedforward networks. Neural Networks 4:251–257

57. Ruder S (2017) An overview of gradient descent optimization algorithms

58. Khirirat S, Feyzmahdavian HR, Johansson M (2017) Mini-batch gradient descent: Faster convergence under data sparsity. In: 2017 IEEE 56th Annual Conference on Decision and Control (CDC). pp 2880–2887

59. YAZAN E, Talu MF (2017) Comparison of the stochastic gradient descent based optimization techniques. In: 2017 International Artificial Intelligence and Data Processing Symposium (IDAP). pp 1–5

60. Smith S, Elsen E, De S (2020) On the Generalization Benefit of Noise in Stochastic Gradient Descent. In: Proceedings of the 37th International Conference on Machine Learning. PMLR, pp 9058–9067

61. Baydin AG, Pearlmutter BA, Radul AA, Siskind JM Automatic Differentiation in Machine Learning: a Survey. 43

62. Güneş Baydin A, Pearlmutter BA, Andreyevich Radul A, Mark Siskind J (2018) Automatic differentiation in machine learning: A survey. Journal of Machine Learning Research 18:1–43

63. Griewank A (2003) A mathematical view of automatic differentiation. Acta Numerica 12:321–398

64. Lu L, Meng X, Mao Z, Karniadakis GE (2021) DeepXDE: A Deep Learning Library for Solving Differential Equations. SIAM Rev 63:208–228. https://doi.org/10.1137/19M1274067

65. Kanghong Li J chai (2003) Comparisons of numerical results from two methods for solving the problem of dam foundation seepage. Hongshui River 22:14–17

66. Mao C (2003) Seepage computation analysis & control. China Hydraulic and Hydropower, Beijing